# Determination of the longitude difference between Baghdad and Khwarezm using a lunar eclipse (the method of Abu Rayhan al-Biruni and Abu al-Wafa al-Buzjani)

R. Bakhromzod[1,2]

[1]S. U. Umarov Physical-Technical Institute, National Academy of Sciences of Tajikistan, Dushanbe 734063, Tajikistan
[2]Institute of Astrophysics, National Academy of Sciences of Tajikistan, Dushanbe 734063, Tajikistan

*Email: rizo@physics.msu.ru

**Abstract.** This paper examines how, in the tenth century, medieval Iranian scholars Abu Rayhan al-Biruni and Abu al-Wafa al-Buzjani determined the difference in geographical longitude between the cities of Baghdad and Khwarezm through simultaneous observation of a lunar eclipse. Brief academic biographies of these scholars are presented, with emphasis on their contributions to mathematics and astronomy. The study discusses the importance of determining geographical coordinates - especially longitude - in the science of the 10th-11th centuries, provides an overview of the methods of coordinate determination available at the time, and highlights the problem of synchronizing remote observations prior to the advent of electronic communication. Particular attention is devoted to a detailed analysis of the method based on observing a lunar eclipse to simultaneously measure longitude differences: the necessary conditions and organization of the experiment, the instruments employed, the mathematical calculations, and error estimates are described. The longitude difference obtained by al-Biruni and al-Buzjani is compared with modern values. The conclusion discusses the scientific significance of this method for the history of science and astronomy.

## Introduction

The determination of accurate geographical coordinates of locations on Earth has been one of the fundamental tasks of Earth sciences (geography and geodesy) since antiquity. In the Middle Ages–especially within Islamic science of the 9th–11th centuries–interest in improving cartographic knowledge and geographical tables increased substantially, which required refining the latitude and longitude of many cities. Knowledge of coordinates was essential both for practical purposes (mapmaking, navigation, and determining the direction toward Mecca–the qibla–in the Islamic world) and for the development of theoretical geography and the verification of data inherited from ancient authors (for example, Claudius Ptolemy). While latitude could be determined relatively easily–by measuring the altitude of the Pole Star or the maximum altitude of the Sun at local noon–the measurement of geographical longitude posed a far greater challenge. Determining the difference in longitude between two locations requires knowledge of the difference in their local times. In medieval conditions, rapid means of communication and sufficiently accurate portable clocks were lacking; consequently, synchronizing time between distant observers constituted a serious scientific and technical problem [1, 2].

This paper examines an outstanding achievement of two tenth-century Persian-Tajik scholars–Abu Rayhan al-Biruni and Abu al-Wafa al-Buzjani–who succeeded in directly measuring the longitude difference between Khwarezm and Baghdad by means of simultaneous observation of a lunar eclipse. This method represented a significant innovation for its time and is one of the few known historical examples of successful longitude determination prior to the advent of modern technologies.

## Abu Rayhan al-Biruni and Abu al-Wafa al-Buzjani: Scholarly Biographies and Their Contributions to Science

### Abu Rayhan al-Biruni (973–1048)

Abu Rayhan Muhammad ibn Ahmad al-Biruni was an Iranian encyclopedic scholar and one of the foremost thinkers of the eleventh century. He was born in 973 in the suburb of Kath (Khwarezm) and acquired the nisba "al-Biruni" ("the outsider" or "from outside the city") from his place of birth. Al-Biruni gained renown for his remarkable works across a wide range of disciplines, including mathematics, astronomy, geography, geodesy, physics, history, and philology (he authored more than one hundred treatises). In the history of science, al-Biruni is often regarded as the "father of geodesy," having made major contributions to determining the size of the Earth and the coordinates of numerous locations. He was among the first in Central Asia to advance ideas compatible with Earth's motion around the Sun (heliocentrism) and proposed an original method for measuring the length of Earth's circumference [3].

Al-Biruni's principal achievements in the mathematical and astronomical sciences include the refinement of observational instruments and methods. He improved the astrolabe and the sextant and constructed a stationary wall-mounted quadrant for precise measurements of celestial positions–at the time the largest instrument of its kind (it remained the largest for nearly four centuries). At the age of twenty-one, al-Biruni determined the obliquity of the ecliptic with remarkable accuracy, attesting to his exceptional observational skills. Later, while in India, he devised an original method for measuring Earth's radius based on the dip of the horizon observed from a mountain summit. By carrying out the relevant observations, al-Biruni computed Earth's circumference with striking precision, very close to the modern value, thereby experimentally confirming the dimensions of the globe [4].

Al-Biruni was deeply engaged in geography and geodesy. He personally determined the coordinates of many cities and regions he visited using astronomical observations (such as measuring the altitudes of celestial bodies). He regarded the determination of geographical coordinates as the geographer's principal task. In his treatise Determination of the Coordinates of Places for the Correction of Distances between Settlements (Kitāb taḥdīd nihāyāt al-amākin li-taṣḥīḥ masāfāt al-masākin), he systematized methods for computing latitude and longitude and compiled a catalogue of coordinates for numerous inhabited places. Al-Biruni sought to improve the accuracy of maps and intercity distances relative to Ptolemy's data. In his

works–most notably the celebrated astronomical compendium The Masʿudi Canon–he synthesized accumulated astronomical knowledge, reported the results of dozens of his own observations (including equinoxes, eclipses, and planetary transits), and presented methods for calculating the positions of celestial bodies. The Masʿudi Canon, dedicated to Sultan Masʿud of Ghazna in 1030, is an extensive encyclopedia of contemporary astronomy and astrology, containing astronomical tables, coordinates of stars and cities, and descriptions of instruments and observational techniques; it long remained an authoritative reference for later scholars [5, 6, 7].

Al-Biruni also demonstrated notable talent in mathematics. In his treatise The Book on the Determination of Chords in a Circle, he presented original solutions and proofs, including reductions of the problems of angle trisection and cube duplication to the solution of cubic equations. Although these achievements pertain to pure mathematics, they reflect a high level of mastery of geometry–an essential foundation for astronomical computation. Al-Biruni maintained correspondence and engaged in scholarly debates with other leading thinkers of his era; for example, he discussed problems of physics and cosmology with the Persian-Tajik polymath Ibn Sina (Avicenna). As a versatile scholar, al-Biruni remained committed to an experimental approach and critically reassessed established doctrines when contradicted by observation. His work in geodesy and astronomy related to longitude determination–examined in detail below–stands as a vivid example of this methodological stance [8].

### Abu al-Wafa al-Buzjani (940–998)

Abu al-Wafa al-Buzjani (Abu al-Wafa Muhammad ibn Muhammad al-Buzjani) was a tenth-century Iranian astronomer and mathematician, a contemporary from the older generation relative to al-Biruni. He was born in 940 in the city of Buzjan (Khorasan, in present-day Iran) and died in 998 in Baghdad. Abu al-Wafa gained renown as one of the greatest mathematicians of the medieval East, making seminal contributions to the development of trigonometry and astronomical computation. He was the first to systematize trigonometry as an independent discipline: he introduced the concepts of the tangent and cotangent (designated in Arabic manuscripts as al-ẓill and quṭr al-ẓill, respectively) and computed tables of their values. He also calculated sin 1° with high precision and derived the formula for the sine of a sum, $\sin(\alpha+\beta)$, thereby substantially simplifying astronomical calculations. Together with another eminent Tajik mathematician, al-Khujandi, he proved the sine theorem for spherical triangles–an essential result of spherical trigonometry required for celestial astronomy. These achievements markedly improved the accuracy of astronomical tables and computations of the period [9, 10].

Abu al-Wafa al-Buzjani also made important contributions to practical astronomy. From the 960s onward he worked at the court of the Buyid emirs in Baghdad and conducted regular observations of celestial phenomena. In 988, at the behest of the Baghdad ruler Sharaf al-Dawla, he founded an astronomical observatory in Baghdad equipped with large instruments. According to historical accounts, it housed a massive wall-mounted quadrant with a diameter of approximately six meters (25 cubits) for high-precision measurements of celestial positions. The existence of such an observatory attests to the exceptionally high level of astronomical practice achieved by Abu al-Wafa and his colleagues.

Like al-Biruni, Abu al-Wafa combined theoretical work with systematic observation. He authored a revised exposition of Ptolemy's Almagest (an alternative presentation and correction of Ptolemy's treatise), in which he synthesized established astronomical data and supplemented them with the results of his own observations, including the lunar inequality noted above. He also compiled extensive astronomical tables–the Comprehensive Zīj (al-Zīj al-Shāmil)–containing coordinates of stars and planets, eclipse schedules, and other information necessary for calendrical computations and the prediction of celestial phenomena. These tables and manuals (zījes) remained in use by astronomers for many years after his death. In addition, Abu al-Wafa wrote a number of treatises on geometry and mathematics, including commentaries on the works of major predecessors–al-Khwarizmi, Euclid, Diophantus, and Hipparchus–demonstrating a profound command of geometric constructions [11, 12].

## The Significance of Coordinate Determination in Medieval Science

In the Middle Ages–particularly during the Islamic "Golden Age" of science–accurate determination of the geographical coordinates of cities and regions was of paramount importance for cartography, navigation, and astronomy. Coordinates–latitude and longitude–made it possible to situate locations within a global reference framework and thereby construct reliable world maps. Latitude (the angular distance from the equator) could be measured relatively easily by astronomical observations: it sufficed to determine the altitude of the Pole Star above the horizon at night or the Sun's maximum altitude at local noon, quantities directly related to a site's latitude. Already in antiquity, geographers and astronomers (Eratosthenes, Ptolemy, and others) had accumulated methods and results for determining the latitudes of many locations [13].

Longitude–the angular distance along a parallel from a chosen prime meridian–posed a far greater challenge. In principle, longitude can be determined by knowing the time difference between a given location and a reference meridian (for example, Greenwich in the modern system or the "Fortunate Isles" in Ptolemy's geography). Ideally, this requires comparing simultaneous clock readings at two locations; however, in the Middle Ages there were neither adequate means of communication nor sufficiently accurate timekeeping devices to perform such a procedure. An error of even a few minutes translated into errors of several degrees in longitude (since one hour corresponds to 15° of longitude). Consequently, longitude determination constituted a major scientific problem.

Nevertheless, medieval Tajik scholars clearly recognized the importance of the task. Accurate coordinates were needed not only for cartography but also for astronomical applications (such as compiling star charts and calculating the visibility of eclipses and planetary conjunctions in different regions), as well as for calendrical and time-reckoning purposes. In Islamic astronomy in particular, determining longitude differences between observatories and cities was crucial, as it enabled the creation of unified zījes (astronomical tables) and the comparison of observations made at different locations by reducing them to a common time reference. Moreover, Islamic geography had a practical motivation: determining the qibla, the direction toward Mecca for any given location, which required knowledge of longitude and latitude differences between that place and Mecca. For these reasons, geodetic expeditions and coordinate-determination observations were actively encouraged by rulers and patrons of science [14, 15].

## Review of Methods for Determining Coordinates Prior to the Eleventh Century

The determination of latitude in antiquity and the Middle Ages was based on measurements of the altitudes of celestial bodies. Observers employed relatively simple astronomical instruments, including gnomons (vertical rods used to measure the length of a shadow at noon), astrolabes, and quadrants. Latitude φ is related either to the Sun's maximum altitude at local noon on the day of the equinox, $h_{max} = 90° − φ$, or to the altitude of the Pole Star above the horizon, $h_{polaris} ≈ φ$ [16, 17].

These techniques were known since Ancient Greece and were widely used by Muslim astronomers. Determining longitude, however, required establishing a difference in time. Ancient geographers often relied on approximate longitude estimates derived from travel distances and assumed speeds of movement. In his *Geography*, Ptolemy listed longitudes for cities, but many of these values contained substantial errors (up to several tens of degrees), since no precise method for measuring longitude was available to him. In some cases, ancient astronomers attempted to use simultaneous observations. It is known, for example, that Hipparchus compared records of a lunar eclipse observed in Sicily and Alexandria and attempted to compute the longitude difference between these locations from the reported time discrepancy. Such data, however, were rare and not always reliable [13, 18].

By the early Middle Ages (9th–10th centuries), the Islamic world had accumulated astronomical tables (*zījes*) that enabled the prediction of the timing of eclipses and other events with a certain degree of accuracy. This created a technical possibility to know in advance when, for instance, a lunar eclipse would occur and to organize observations in different cities. In addition, time-measuring instruments were improved, including nocturnal water clocks (clepsydrae), stellar clocks, and astrolabes. Although the accuracy of timekeeping remained limited (on the order of several minutes), these developments opened the way to implementing ancient ideas at a new level.

Prior to this period, only isolated attempts had been made to exploit such an approach. For example, another Iranian scholar, al-Khwarizmi (9th century), and other compilers of astronomical tables included data on eclipse times, but no systematic program of synchronized observations was undertaken. In the tenth century, major observatories operated in different regions of the Islamic world (for example, the observatory of al-Khujandi in Rayy, and the Cairo observatory associated with Ibn Yunus), yet each functioned autonomously [10]. The idea of combining observations from two observatories to measure a geographical parameter represented a novel step forward.

### The Problem of Synchronizing Distant Observations (10th–11th Centuries)

To implement simultaneous observations in the 10th–11th centuries, scholars had to overcome a fundamental difficulty: the absence of instantaneous communication. It was impossible to transmit a real-time signal from Baghdad to Khwarezm, or vice versa, indicating the occurrence of a specific moment. Any attempt to use messengers, signal fires, or sounds was constrained by the limited speed of information propagation and by severe distortions over long distances [19]. Consequently, scholars realized that the only reliable "time signal" simultaneously accessible in different locations was a natural phenomenon observable at the same moment.

In an astronomical context, such phenomena could include the outburst of a new star, the appearance of a comet, or an occultation (eclipse) of a celestial body. Stellar outbursts and comets, however, are unpredictable, and lunar occultations of stars are visible only over limited regions. A lunar eclipse, by contrast, is ideally suited for this purpose: when the Moon enters Earth's shadow, the event is visible across the entire nighttime hemisphere–from Morocco to China. It is therefore unsurprising that lunar eclipses came to be regarded as a kind of "calendrical marker" for synchronization [20].

Even when observing a single event, another problem remained: how to record the exact moment locally at each site. In the tenth century, daily time was determined locally using sundials (by day) or water clocks; at night, clepsydrae and stellar observations were employed. Astronomers were able to determine local time from the positions of known stars using an astrolabe. Nevertheless, the precision of such timekeeping was limited. Accordingly, observers agreed in advance on which specific phase of the eclipse would be compared–for example, the beginning of totality or the midpoint of the eclipse–and attempted to measure the duration from a reference moment (such as sunset) to that phase.

Thus, the synchronization problem was reduced to the following tasks: agreeing in advance on which phenomenon would be observed and which moment of it would be recorded on each observer's timekeeping device; ensuring that the phenomenon would indeed be visible at both locations (for instance, that the Moon would be above the horizon simultaneously in Khwarezm and Baghdad during the eclipse); and finally, accounting for potential timing errors–differences in clock rates, non-instantaneous human reaction times, and similar factors–and minimizing their effects.

Al-Biruni and Abu al-Wafa al-Buzjani approached these challenges with particular care. They used correspondence to coordinate their actions. By computing in advance, from astronomical tables, the date and approximate time of an upcoming lunar eclipse, they agreed to conduct simultaneous observations and subsequently exchange their results. In this way, the astronomical phenomenon itself served as the synchronizing "signal," while written correspondence functioned as the carrier of information about the time difference after the experiment.

It should be emphasized that such coordination was highly innovative for the tenth century. In effect, al-Biruni and al-Buzjani organized a form of "simultaneous observation session" over a distance of approximately 1,500 kilometers, without any direct contact during the experiment itself. This required a high degree of confidence in the precomputed predictions (so that both observers knew when to expect the eclipse) and trust in their timekeeping instruments and observational methodology.

**The Lunar Eclipse Method for Measuring Longitude Differences**

## Conditions and Organization of the Observations

The historical experiment to determine the longitude difference between Baghdad and Khwarezm was carried out on 24 May 997 CE (according to calculations, during the night of 24 May in the Julian calendar). At that time, the 24-year-old Abu Rayhan al-Biruni was in his native Khwarezm (the city of Kath, the capital of Khwarezm) after a period of study and travels, while the 57-year-old Abu al-Wafa al-Buzjani was living and working in Baghdad at the Buyid court. Despite the considerable age difference, the two scholars maintained active scientific correspondence. From al-Biruni's letters it is known that he reached an agreement with al-Buzjani to conduct a joint observation of the forthcoming lunar eclipse: "through correspondence he agreed with the eminent mathematician Abu al-Wafa … on a simultaneous observation of the lunar eclipse from Baghdad and Kath." The observers agreed in advance to record the time of a specific phase of the eclipse–most likely the beginning of the partial phase or the moment of maximum eclipse–using their respective local timekeeping methods.

According to modern calculations (the Espenak–Meeus canon), this eclipse was partial and belonged to Saros series 85. The instant of greatest eclipse occurred at 00:01:22 TD (as given in NASA tables) with $\Delta T = 1577$ s. This implies that, in Universal Time (UT), the maximum took place at approximately 23:35:05 UT on the previous day–that is, during the night of 24/25 May 997 CE–precisely the interval suitable for synchronized observations at both locations [21].

The key geometrical parameters of the eclipse are as follows. The value $\gamma = 0.7145$ indicates that the axis of the Moon's orbit passed significantly offset from the center of Earth's shadow; therefore, the eclipse could not be total. This is confirmed by the umbral magnitude Um. Mag. = 0.5510: at maximum, about 55% of the lunar diameter was immersed in Earth's umbra, producing a clearly visible "bite" out of the lunar disk with a sharp shadow boundary. At the same time, the penumbral magnitude Pen. Mag. = 1.5426 indicates a deep passage through the penumbra: the entire lunar disk lay within the penumbra, with part entering the umbra, resulting in a pronounced darkening readily noticeable even without instruments.

In terms of duration, the event was particularly convenient for precise time recording. The penumbral phase lasted approximately 301 minutes, while the partial (umbral) phase lasted about 157 minutes. Counting these intervals from the moment of maximum yields approximate phase boundaries (in the TD scale): the penumbral phase began around 21:30 TD on 24 May and ended near 02:32 TD on 25 May; the partial phase extended roughly from 22:43 TD on 24 May to 01:20 TD on 25 May. Converted to UT, this corresponds to approximately 21:04–02:05 UT for the penumbral phase and 22:16–00:53 UT for the partial phase, meaning that the eclipse spanned a substantial portion of the night.

The visibility map (Fig. 1) shows that the observable region encompassed both Mesopotamia and Central Asia: both Baghdad and Khwarezm lay within zones from which the eclipse was visible. In practical terms, this implied the following.

- **In Baghdad** (located farther west and south), the eclipse occurred during deep night: the Moon was well above the horizon, allowing the observer to record the agreed-upon moment–such as the maximum or one of the clearly identifiable contacts of the partial phase (entry into or exit from the umbra), when the shadow edge touches the lunar disk–without haste.

- **In Kath (Khwarezm)**, situated farther east, the same phases were shifted forward in local time toward the pre-dawn hours. The Moon remained observable, but toward the end of the event it approached lower altitudes above the horizon and increasing twilight illumination. Consequently, recording the maximum or the initial contacts of the partial phase–when shadow contrast is highest– was especially valuable for accurate timing.

It was precisely these characteristics–a sufficiently large umbral magnitude (0.55), a long partial phase (157 minutes), and a wide visibility zone–that made the eclipse of 997 CE an ideal "natural time signal" for comparing local times in Baghdad and Khwarezm and, consequently, for estimating the difference in their geographical longitudes.

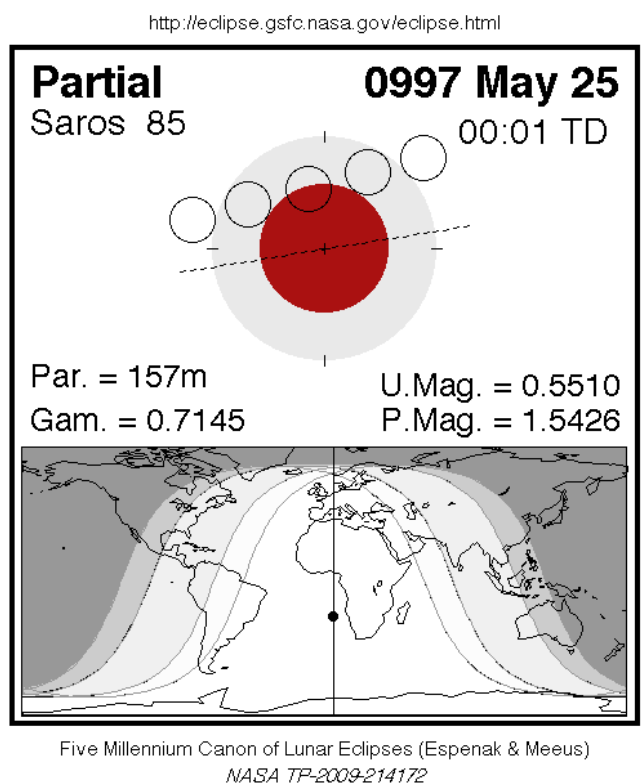

**Figure 1.** Partial lunar eclipse of 25 May 997 CE: phase diagram, principal parameters, and visibility map (Saros 85; Um. Mag. = 0.551; γ = 0.7145).

The choice of a lunar eclipse was optimal. First, such an eclipse is guaranteed to be observable simultaneously in both regions, provided that the Moon is above the horizon. On 24 May 997 CE, the Moon indeed rose in the evening both in Baghdad and in Kath (Khwarezm), and the eclipse occurred during the first half of the night. Second, a lunar eclipse is sufficiently long-lasting (the total phase may last up to about an hour), allowing astronomers ample time to record the required moment and check their timekeeping. Third, unlike stellar occultations, a lunar eclipse does not depend on the observer's longitude or latitude, since Earth's shadow covers the Moon simultaneously for all observers on the nighttime hemisphere. Thus, the condition of simultaneity required for the observations was satisfied by the very nature of the phenomenon.

When the eclipse occurred, each scholar recorded the local time of a key phase. It is known, for example, that Abu Rayhan al-Biruni in Kath noted the hour of the night at which the decisive phase began (most likely the moment of greatest eclipse or the entry into the umbra). Similarly, Abu al-Wafa al-Buzjani recorded the time of the same phase according to Baghdad local time. Naturally, there was no direct communication between the observers during the event; each worked autonomously. After the eclipse, the results had to be compared. To this end, it was most likely arranged in advance that the data would be exchanged by letter. Either al-Biruni sent his report to Baghdad or, conversely, the senior colleague al-Buzjani transmitted his record to Khwarezm. Several weeks later, when a caravan or courier delivered the letter, the scholars were able to compare the two time records of the same astronomical event.

It is important to note that each observer needed to verify the reliability of their timekeeping. According to reports, the Baghdad observatory possessed water clocks (clepsydrae) and possibly sandglasses. In Khwarezm, al-Biruni likely had access to a portable clepsydra or relied on stellar observations to determine time. Because the eclipse occurred at night, sundials were not applicable. The astronomers could, for instance, calibrate the water clocks in advance during the day (using noon or stellar observations at sunset) and then maintain a continuous count during the eclipse. In Baghdad, given the presence of a large quadrant and a staff of astronomers, time measurements were probably fairly precise, with uncertainties on the order of minutes. In Khwarezm, al-Biruni, working alone, may have faced greater difficulties in marking time, but his experience and skill evidently allowed him to meet the challenge.

Once the results were finally combined, al-Biruni and al-Buzjani obtained the desired quantity: the difference in local time between Kath and Baghdad at the moment of the eclipse. By comparing their recorded times, they determined that the lunar eclipse occurred in Khwarezm approximately one hour later than in Baghdad (or, equivalently, that the same phase was reached in Baghdad about one hour earlier). In other words, when the Moon had already entered Earth's shadow in Baghdad, the local time in the more easterly Kath lagged by roughly one hour. This time difference constituted the key datum for computing the difference in geographical longitude between the two locations.

### Computation of the Longitude Difference: Formulas and Calculations

Moving from observations to calculations, the scholars relied on a simple relationship between astronomy and geography: over one day (24 hours) the celestial sphere completes a full rotation of 360°. It follows that in 1 hour of rotation the Earth turns by 15° of longitude.

The difference in local time Δt between two locations is related to the difference in their longitudes Δλ by the relation **Δλ = 15° · Δt**, where Δt is expressed in hours (or fractions of an hour).

In the case of Baghdad and Khwarezm (Kath), the difference in local time determined by Abu Rayhan al-Biruni and Abu al-Wafa al-Buzjani amounted to approximately 1 hour. Substituting Δt ≈ 1 h yields Δλ ≈ 15°. Thus, according to their measurements, the meridian of Khwarezm lay about 15° east of the meridian of Baghdad (equivalent to a longitude difference of approximately 15°).

In his notes, al-Biruni remarked that the difference between the local noons (and hence the longitude difference) of Kath and Baghdad was close to one hour. In his later works, he explicitly recorded this longitude difference. For example, in his geographical tables he gave the longitude of Kath (Khwarezm) with a correction of 1 hour relative to Baghdad. Unfortunately, the exact numerical values originally computed by the scholars have not survived in their original form; they may have expressed longitudes relative to an adopted prime meridian. In Islamic geography, the prime meridian was often taken to be that of the "Fortunate Isles" (as in Ptolemy) or the western boundary of the inhabited world. If Baghdad was assigned a longitude of about 79° from Ptolemy's reference meridian, then adding 15° would yield approximately 94° for Khwarezm. For the present analysis, however, the relative difference is of primary importance.

It can thus be reconstructed that al-Biruni and al-Buzjani obtained a result on the order of 15° for the eastward longitude difference. Al-Biruni later emphasized this experiment with pride as an example of successful longitude determination "by means of simultaneous observation of an eclipse"–a method he applied in practice for the first time within Islamic astronomy.

It should be noted that the calculations may have incorporated not only the midpoint of the eclipse but also other measured parameters, potentially improving accuracy. For instance, if both observers recorded the moments of entry into and exit from the umbra, the midpoint of the eclipse could be determined more precisely as the arithmetic mean of these times and used in the comparison. Moreover, al-Biruni was well versed in eclipse theory and could have employed known astronomical data (e.g., eclipse duration and the Moon's position) to cross-check the result.

In general, the calculations themselves posed little difficulty for scholars of this caliber; the principal challenge lay in accurately determining Δt. Once Δt had been established, it sufficed to multiply it by 15° and record the result using the formula **Δλ = 15° · Δt**. Accordingly, the longitude difference between Baghdad and Khwarezm was reported to be approximately 15°.

### Comparison with Modern Data

What, then, is the actual longitude difference between Baghdad and Khwarezm, and how accurate was the estimate by al-Biruni and al-Buzjani? Modern coordinates allow a quantitative assessment. Baghdad has a longitude of about 44.4009° E, whereas Kath (the historical center of Khwarezm, commonly localized near modern Biruni) lies at approximately 60.750° E. The resulting longitude difference is $\Delta\lambda \approx 16.35°$. Expressed in local time, this corresponds to $\Delta t\_real = \Delta\lambda / 15 \approx 1.0899$ h, that is, about 1 h 05 min 24 s.

According to al-Biruni's account, he and al-Buzjani obtained $\Delta t \approx 1$ h, equivalent to $\Delta\lambda \approx 15°$. Their value is therefore smaller by about 1.35°, corresponding to roughly 5 min 24 s in time. The relative error in the longitude difference is approximately 8.3%, and when expressed as a fraction of Earth's full diurnal rotation it amounts to only about 0.38%. For field measurements of the 11th century, this level of accuracy is truly remarkable: the relative positions of the meridians of Baghdad and Kath were determined correctly in order of magnitude and fairly closely in absolute terms.

In terms of distance, 1° of longitude at these latitudes corresponds to roughly 85–90 km (more precisely, about 88 km at a mean latitude of ~37–38°). Consequently, an error of $\delta\lambda \approx 1.35°$ implies an east–west displacement of about $\delta S \approx 120$ km. Against the backdrop of the total distance between the regions of Baghdad and Khwarezm–exceeding 1,500–1,700 km–this error amounts to roughly 7–8%. For medieval geography, where distances were often estimated with accuracies on the order of 10–15%, this represents an exceptionally good result.

## Conclusion

The experiment conducted by Abu Rayhan al-Biruni and Abu al-Wafa al-Buzjani to determine the longitude difference between Khwarezm and Baghdad through observation of a lunar eclipse is rightly regarded as an outstanding achievement of medieval science. It represents a remarkable synthesis of theoretical insight and practical implementation, astronomical knowledge and geographical purpose. The scholars succeeded in overcoming the principal limitation of their era–the absence of means for time synchronization–by using the sky itself as a set of "precise clocks."

In summary, the method of determining longitude differences by means of a lunar eclipse, implemented by tenth-century Iranian astronomers, constitutes a vivid chapter in the history of science. The experience of al-Biruni and al-Buzjani demonstrates how the high scientific culture of the medieval East, building upon classical heritage and its own innovations, achieved results that remain worthy of admiration today.

**Note.** An AI-based language model (ChatGPT, version 5.2) was used to assist in the English translation and linguistic polishing of the manuscript; the authors take full responsibility for the content.

## References

1. Brauer, R. W. *Boundaries and Frontiers in Medieval Muslim Geography*. American Philosophical Society, 1995, vol. 85.
2. Schoy, C. "The Geography of the Moslems of the Middle Ages." *Geographical Review*, 1924, vol. 14, no. 2, pp. 257–269.

3. Kamiar, M. *A Bio-Bibliography for Biruni: Abu Raihan Mohammad Ibn Ahmad (973–1053 CE)*. Bloomsbury Publishing PLC, 2006.
4. Maslikov, S. Yu. "How al-Biruni Measured the Earth." *Geodesy and Cartography*, 2019, vol. 80, no. 7, pp. 57–64. (in Russian)
5. Ahmed, A. "Al-Biruni: One of the Greatest Pioneers of Geographical Science." *VFAST Transactions on Islamic Research*, 2022, vol. 10, no. 3, pp. 18–22.
6. al-Biruni, A. R. *Selected Works*, vol. III: *Geodesy*. Edited, translated, and annotated by P. G. Bulgakov. Tashkent: Fan, 1963. (in Russian)
7. Boboev, Yu. A., and Mansurova, M. A. "Fixing the Longitudes and Latitudes of Cities in Tables (Translation of Chapter 10, Book V of *The Masʿudi Canon* by Abu Rayhan al-Biruni into Tajik)." *Bulletin of the Tajik State University of Law, Business and Politics. Humanities Series*, 2024, vol. 100, no. 3, pp. 5–17.
8. Gafurov, B. "Al-Biruni: A Universal Genius Who Lived in Central Asia a Thousand Years Ago." *The UNESCO Courier*, 1974, vol. XXVII, no. 6, pp. 4–9.
9. Starr, S. F. *Lost Enlightenment: Central Asia's Golden Age from the Arab Conquest to Tamerlane*. Moscow: Alpina Publisher, 2023. (Russian edition)
10. Abdullazade, Kh. F., and Negmatov, N. N. *Abu Mahmud al-Khujandi*. Dushanbe, 1986. (in Russian)
11. Neugebauer, O., and Rashed, R. "Sur une construction du miroir parabolique par Abū al-Wafāʾ al-Būzjānī." *Arabic Sciences and Philosophy*, 1999, vol. 9, no. 2, pp. 261–277.
12. Kolman, E., and Yushkevich, A. P. *Mathematics before the Renaissance*. Moscow: Fizmatgiz, 1961. (in Russian)
13. Wright, J. K. "Notes on the Knowledge of Latitudes and Longitudes in the Middle Ages." *Isis*, 1923, vol. 5, no. 1, pp. 75–98.
14. King, D. A., Samsó, J., and Goldstein, B. R. "Astronomical Handbooks and Tables from the Islamic World (750–1900): An Interim Report." *Suhayl: International Journal for the History of the Exact and Natural Sciences in Islamic Civilisation*, 2001, pp. 9–105.
15. Abdullah, D. "Determining the Qibla Direction by Astronomical and Geometrical Methods." *arXiv* preprint arXiv:2512.03271, 2025.
16. Sparavigna, A. C. "Measuring Times to Determine Positions." *arXiv* preprint arXiv:1202.2746, 2012.
17. Knox-Johnston, R. "Practical Assessment of the Accuracy of the Astrolabe." *The Mariner's Mirror*, 2013, vol. 99, no. 1, pp. 67–71.
18. Russo, L. "Ptolemy's Longitudes and Eratosthenes' Measurement of the Earth's Circumference." *Mathematics and Mechanics of Complex Systems*, 2013, vol. 1, no. 1, pp. 67–79.
19. Chiat, B. "Longitude Determination: An Historical Survey." *Monthly Notes of the Astronomical Society of South Africa (MNASSA)*, 1956, vol. 15, no. 11, pp. 91–95.
20. Portuondo, M. M. "Lunar Eclipses, Longitude and the New World." *Journal for the History of Astronomy*, 2009, vol. 40, no. 3, pp. 249–276.
21. *Five Millennium Catalog of Lunar Eclipses*. NASA Goddard Space Flight Center. https://eclipse.gsfc.nasa.gov/LEcat5/LE0901-1000.html